\documentclass[a4paper,aps,11pt,nofootinbib,showpacs]{revtex4}

\usepackage{amsmath}
\usepackage{amssymb}
\usepackage{amsfonts}
\usepackage[dvips]{color,graphicx}
\usepackage{subfigure}

\begin{document}
\title{{\Large Stable Bound Orbits of Massless Particles around a Black Ring}}

\hfill{OCU-PHYS 376}

\hfill{AP-GR 103}


\author{$^{1,2}$Takahisa Igata} 
\email{igata@sci.osaka-cu.ac.jp}
\author{$^{2}$Hideki Ishihara}
\email{ishihara@sci.osaka-cu.ac.jp}
\author{$^{2}$Yohsuke Takamori}
\email{takamori@sci.osaka-cu.ac.jp}
\affiliation{%
$^{1}$Department of Physics, Kinki University, Osaka 577-8502, Japan\\
$^{2}$Department of Mathematics and Physics,
 Graduate School of Science, Osaka City University,
 Osaka 558-8585, Japan}

\begin{abstract}
We study the geodesic motion of massless particles in singly rotating black ring 
spacetimes. We find stable stationary orbits of massless particles in toroidal spiral 
shape in the case that the thickness parameter of a black ring is less than a 
critical value. Furthermore, there exist nonstationary massless particles bounded in 
a finite region outside the horizon. This is the first example of stable bound orbits 
of massless particles around a black object.
\end{abstract}

\maketitle

\section{Introduction}
Higher-dimensional black holes gather much attention in the context of modern unified 
theories of interactions (see for a review~\cite{LivingReview}). It is understood that the 
higher-dimensional black holes are very different from the four-dimensional ones 
in some aspects. In four dimensions, stationary black holes in a vacuum must have 
spherical horizons, and are uniquely characterized by their mass and angular momentum. 
In higher dimensions, however, black holes with unusual horizon topology exist 
in addition to black holes with spherical horizon topology~\cite{Myers:1986un}. The first 
discovery of the black hole solutions with ${\rm S}^2\times{\rm S}^1$ 
topology, black ring solutions, was made in five dimensions by Emparan and 
Reall~\cite{Emparan:2001wn}. The black ring solutions reveal that a black hole in a vacuum 
is not uniquely characterized only by its mass and angular momenta in higher dimensions. 
After this pioneering work, various types of black ring solutions have been found 
by many authors~\cite{Pomeransky:2006bd, Mishima:2005id, Elvang:2007rd, Iguchi:2007is, 
Evslin:2007fv, Izumi:2007qx, Elvang:2007hs}. It is important to study their geometrical 
properties and physical phenomena that occur around each of them in order to 
understand the differences between black holes and black rings.

One of the important steps to clarify the geometry of spherical black holes and black 
rings is the studies of geodesics in these spacetimes. For example, the separability in the 
Hamilton-Jacobi equation for geodesic motion, which is a general property 
in arbitrary-dimensional Myers-Perry black holes~\cite{Frolov:2003en}, implies that the black 
holes have the hidden symmetries called Killing tensors. On the other hand, in the case of 
singly rotating black rings, such separation of variables occurs for some special cases of 
geodesics: the geodesics on the rotational axis or the equatorial plane, and the geodesics 
for the massless particles with zero-energy~\cite{Hoskisson:2007zk, Durkee:2008an}. 
Recently, the Hamilton-Jacobi equation in these separable cases were solved 
analytically~\cite{Grunau:2012ai, Grunau:2012ri}.

In four dimensions, a massive particle moving around a black hole has two types of orbits: 
bound orbits by the gravitational force, and unbound ones. A part of bound orbits are 
bounded in a region outside the black hole horizon, where the centrifugal force prevents the particles 
from falling into the horizon. There are circular orbits that are stable against small 
perturbations as special cases of the bounded orbits. In contrast, there is no stable circular 
orbit of massive particles in higher-dimensional Schwarzschild black holes, nor at least on 
the equatorial planes in the five-dimensional Myers-Perry black holes~\cite{Frolov:2003en}. 
This fact is completely different from the four-dimensional case. However, stable bound 
orbits of massive particles exist around a black ring if the thickness parameter is less than a critical 
value~\cite{Igata:2010ye}. The result suggests that we can distinguish a black ring from a 
spherical black hole with the same mass and the same angular momentum by the orbits 
of massive particles. Furthermore, it is shown that chaotic motion of massive particles 
appears~\cite{Igata:2010cd}. It indicates that there is not the sufficient number of 
constants of motion for integration of geodesic equations for a massive particle 
in black ring spacetimes.

How about massless particles? In four-dimensional black holes, there exists no bounded 
orbit of massless particles in a finite region outside the horizon though there are unstable 
circular orbits of massless particles. Higher-dimensional spherical black holes 
with the asymptotic flatness would have the same property as the four-dimensional ones. 
In contrast, in black holes with the asymptotic structure of the Kaluza-Klein 
type~\cite{Dobiasch:1981vh, Gibbons:1985ac, Ishihara:2005dp}, 
a massless particle with nonvanishing momenta 
in the directions of extra dimensions could be bounded in a finite region. 
The most simple example of such nonasymptotically flat black holes is a black string with 
periodic identification along the string direction. The geometry is given by the direct 
product of a Schwarzschild spacetime with a one-dimensional circle, S$^1$. 
A massless particle that has nonvanishing momentum along S$^1$ behaves as a massive 
particle in the four-dimensional Schwarzschild spacetime. Hence, such a massless particle 
can be bounded around the black string.

Are there bounded orbits of massless particles around a black ring? On the one hand, 
since a black ring spacetime is asymptotically flat, it seems that massless particles cannot 
be bounded around it. On the other hand, since the geometry of a black ring approaches the 
one of a black string in a region near the horizon, then we expect that a massless particle 
could be bounded around it if the particle had nonvanishing angular momenta in two 
independent angular directions. 

In this paper, we investigate null geodesics in the singly rotating black ring geometry in five
 dimensions as a moving particle system in a two-dimensional potential. We show the 
 existence of stable bound orbits of massless particles and the existence of bounded 
 orbits of massless particles generally in a finite region outside the horizon.

The paper is organized as follows. The following section provides the singly rotating black 
ring geometry and the Hamilton formalism for a massless particle around a black ring. 
In Sec.~\ref{sec:3}, we show the existence of stable bound orbits of massless particles 
in the case that the black ring thickness parameter is smaller than a critical value. 
Section~\ref{sec:4} shows that there exist more general orbits of massless particles bounded 
in a finite domain outside the black ring horizon.

\section{Hamiltonian for a massless particle}
\subsection{Geometry of singly rotating black rings}

We consider the metric of singly rotating black rings given by
\begin{align}
	ds^2 =& g_{\alpha\beta} dx^\alpha dx^\beta 
\cr
	=& -\frac{F(y)}{F(x)}\left(dt-CR\frac{1+y}{F(y)}d\psi\right)^2
\cr&
	+\frac{R^2}{(x-y)^2}F(x)
	\left(-\frac{G(y)}{F(y)}d\psi^2 - \frac{dy^2}{G(y)} + \frac{dx^2}{G(x)} + \frac{G(x)}{F(x)}d\phi^2\right),
\label{eq:metric}
\end{align}
with
\begin{align}
	&F(\xi) = 1+\lambda\xi,
\quad
	G(\xi) = (1-\xi^2)(1+\nu \xi),
\\
	&C = \sqrt{\lambda(\lambda-\nu)\frac{1+\lambda}{1-\lambda}},
\end{align}
where the ranges of the ring coordinates $x$ and $y$ are given by 
\begin{align}
	-1 \leq x \leq 1~~ {\rm and}~ -\infty \leq y \leq -1,
\end{align}
and the angular coordinates $\phi$ and $\psi$ are periodic in $2\pi$. 
The parameter $R$ denotes the radius of the black ring. The parameters $\nu$ and $\lambda$ 
describe the thickness of the black ring and the rotational velocity in the $\psi$-direction 
through $C$, respectively. These parameters should be chosen in the range
\begin{align}
	0<\nu\leq\lambda<1, 
\end{align}
and related as
\begin{equation}
	\lambda = \frac{2\nu}{1+\nu^2}
\end{equation}
to avoid the conical singularities at $x=\pm1$ and $y=-1$. 
The event horizon and the ergosurface in the ${\rm S}^2\times {\rm S}^1$ topology 
are located at $y=-1/\nu$ and $y=-1/\lambda$, respectively.

\subsection{Hamiltonian formalism}
Throughout this paper, we use the Hamiltonian formalism for a massless particle. Motion of a 
massless particle is governed by the Hamiltonian,
\begin{equation}
	H = \frac{1}{2}g^{\alpha\beta}k_{\alpha}k_{\beta},
\label{eq:H}
\end{equation}
with the null condition
\begin{equation}
	g^{\alpha\beta}k_\alpha k_\beta = 0,
\label{eq:null}
\end{equation}
where $k_{\alpha}$ is the canonical momentum. For the black rings, $g^{\alpha\beta}$ is given 
by the inverse of Eq.~\eqref{eq:metric}.

Since the metric~\eqref{eq:metric} admits three Killing vectors, $\partial_t$, 
$\partial_\psi$, and $\partial_\phi$, there exist three constants of motion,
\begin{equation}
	E = -k_t, ~ L_\psi = k_\psi, ~{\rm and}~ L_\phi = k_\phi,
\label{eq:constants}
\end{equation}
where $E$, $L_\psi$, and $L_\phi$ are energy, angular momenta in the $\psi$- and the 
$\phi$-direction of a massless particle, respectively. Substituting 
Eqs.~\eqref{eq:constants} into Eq.~\eqref{eq:H}, we obtain the two-dimensional effective 
Hamiltonian,
\begin{equation}
	H = \frac{1}{2}\left(g^{xx}k_{x}^2 + g^{yy}k_{y}^2 + E^2\, U\right),
\label{eq:Heff}
\end{equation}
where $U$ is the effective potential given by
\begin{equation}
	U = g^{tt} + g^{\phi\phi}l_{\phi}^2 + g^{\psi\psi}l_{\psi}^2 - 2g^{t\psi}l_{\psi},
\label{eq:U}
\end{equation}
with
\begin{align}
	g^{tt} &= -\frac{F(x)}{F(y)}-\frac{C^2(x-y)^2(y+1)^2}{G(y)F(x)F(y)},
\quad
	g^{xx} = \frac{(x-y)^2}{R^2}\frac{G(x)}{F(x)},
\quad
	g^{yy} = - \frac{(x-y)^2}{R^2}\frac{G(y)}{F(x)},
\cr
	g^{\phi\phi} &= \frac{(x-y)^2}{R^2 G(x)},
\quad
	g^{\psi\psi} = -\frac{F(y)(x-y)^2}{R^2 G(y)F(x)},
\quad
	g^{t\psi} = -\frac{C(x-y)^2(y+1)}{RG(y)F(x)}.
\end{align}
In Eq.~\eqref{eq:U}, we have introduced the normalized angular momenta, $l_\phi := L_\phi/E$ and $l_{\psi} := L_\psi/E$.

The equations of motion for Eq.~\eqref{eq:Heff} are
\begin{align}
 	\dot{x}^i &= g^{ij}k_j, \quad (i, j = x, y),
\label{eq:eom_x}\\
	\dot{k}_i &= - \frac{1}{2}(\partial_i g^{jk}k_j k_k + \partial_i U),
\label{eq:eom_k}
\end{align}
where the overdot denotes the differentiation with respect to an affine parameter 
on each world line of the massless particle. The null condition~\eqref{eq:null} 
becomes
\begin{equation}
	g^{xx}k_{x}^2 + g^{yy}k_{y}^2 + E^2\,U = 0.
\label{eq:null2}
\end{equation}
Note that the effective potential~\eqref{eq:U} is symmetric with respect to $l_\phi \to - l_\phi$, 
whereas asymmetric with respect to $l_\psi \to - l_\psi$, because the black ring described by 
the metric~\eqref{eq:metric} rotates only in the $\psi$-direction. Therefore, we assume 
$l_\phi \geq 0$ without loss of generality in what follows. Further, we consider future-directed 
null geodesics with positive energy $E>0$.

\section{Toroidal spiral orbits}
\label{sec:3}
In this section, we discuss the existence of local minimum points of the effective potential 
$U$. In the case $U=0$ at the points, they correspond to stable stationary orbits 
of massless particles.

\begin{figure}[!t]
\setlength\abovecaptionskip{0pt}
\begin{center}
\includegraphics[width=80mm]{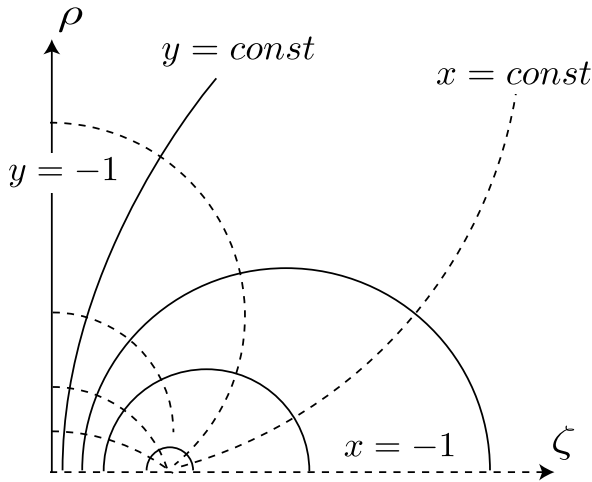}
\end{center}
\caption{\footnotesize
Relation of the $(x,y)$ coordinates and the $(\zeta, \rho)$ coordinates. The solid curves 
denote $y={\rm const}$, and the dashed curves denote $x={\rm const}$ in the $\zeta$-$\rho$ plane.}
\label{fig:coordinates}
\bigskip
\end{figure}

Throughout the following discussion we use the new coordinates $(\zeta, \rho)$ defined by
\begin{equation}
	\zeta = \frac{R\sqrt{y^2-1}}{x-y} ~~{\rm and}~~ \rho = \frac{R\sqrt{1-x^2}}{x-y}.
\end{equation}
Figure~\ref{fig:coordinates} shows the relations of $(x, y)$ and $(\zeta, \rho)$. 
The new coordinate system has an advantage because the metric is written 
in the well known form,
$ds^2 = - dt^2 + d\zeta^2 +\zeta^2 d\psi^2 + d\rho^2 +\rho^2 d\phi^2$, in the flat limit of 
Eq.~\eqref{eq:metric}. In what follows we employ the unit $R=1$ for simplicity.

We consider the special stationary motion that satisfies
$\dot\zeta = \dot\rho = 0$ and $\dot{k}_\zeta = \dot{k}_\rho = 0$,
or equivalently,
$\dot x = \dot y = 0$ and $\dot{k}_x = \dot{k}_y = 0$.
Then the equations of motion~\eqref{eq:eom_x} and \eqref{eq:eom_k} lead to $k_\zeta=k_\rho=0$ 
and the stationary point conditions for $U$,
\begin{equation}
	\partial_\zeta U = \partial_\rho U = 0,
\label{eq:dU=0}
\end{equation}
where $U$ is the function of $\zeta$ and $\rho$ given by Eq.~\eqref{eq:U} as
$U = U\left(x(\zeta, \rho), y(\zeta, \rho)\right)$.
Furthermore, from the null condition~\eqref{eq:null2}, $U$ must satisfy
\begin{equation}
	U=0.
\label{eq:U=0}
\end{equation}

The stationary orbits of massless particles satisfying Eqs.~\eqref{eq:dU=0} 
and \eqref{eq:U=0} with nonvanishing $E$, $l_\psi$, and $l_\phi$, are tangent to 
the null Killing vectors that are linear combinations of $\partial_t$, 
$\partial_\phi$, and $\partial_\psi$. Hence, the projection of the orbit on a $t={\rm const}$ 
surface makes a toroidal spiral curve on a two-dimensional torus, a direct product of S$^1$ 
with a constant radius $\zeta$ generated by $\partial_\psi$ and S$^1$ with a constant radius 
$\rho$ generated by $\partial_\phi$. Then, we call such a stationary orbit 
a {\it toroidal spiral orbit} of a massless particle.

\subsection{Set of Stationary points}
Since $U$ includes the two parameters $l_\phi$ and $l_\psi$, then it is useful to consider 
the direct product space: the two-dimensional space of $(\zeta, \rho)$ times the 
two-dimensional parameter space of $(l_\psi, l_\phi)$, i.e., the four-dimensional space 
defined as ${\cal N}=\{(\zeta, \rho, l_\psi, l_\phi)\}$. In $\cal N$, Eqs.~\eqref{eq:dU=0} 
defines a two-dimensional surface, say $\Sigma$, which represents the set of stationary points.

We consider the restriction of $U$ to $\Sigma$, denoted by $U|_\Sigma$. The condition
\begin{equation}
	U|_\Sigma=0
\label{eq:null_rest}
\end{equation}
makes a curve on $\Sigma$, and each point on the curve corresponds to a stationary orbit 
of a massless particle of $\zeta = {\rm const}$ and $\rho = {\rm const}$, i.e., a stationary 
toroidal spiral orbit. In particular, if a point on $\Sigma$ satisfying the 
condition~\eqref{eq:null_rest} is a local minimum of $U$, the corresponding stationary toroidal 
spiral orbit is stable. The condition for a stationary point of $U$ to be a local minimum is 
\begin{equation}
	\left.\det{\cal H}(U)\right|_\Sigma > 0 ~~{\rm and}~~ 
	\left.{\rm tr}\,{\cal H}(U)\right|_\Sigma > 0
\label{eq:stability}
\end{equation}
at the point, where ${\cal H}(U)|_\Sigma$ is the restriction of the Hessian matrix of $U$,
\begin{align}
	{\cal H}(U) = \left(
	\begin{array}{ccc}
		\partial_\zeta^2U & \partial_\zeta\partial_\rho U\\
		\partial_\rho \partial_\zeta U & \partial_\rho ^2U\\
	\end{array}
	\right),
\end{align}
to $\Sigma$. In order to seek local minimum points of $U$ satisfying Eq.~\eqref{eq:null_rest}, 
we find $\Sigma$ first, and then we inspect the conditions~\eqref{eq:null_rest} and 
\eqref{eq:stability}.

Here, we study the surface $\Sigma$ in detail. Equations~\eqref{eq:dU=0} yield the quadratic 
equation in $l_\psi$, 
\begin{equation}
	\alpha l_\psi^2 + 2\beta l_\psi + \gamma = 0,
\label{eq:lpsi}
\end{equation}
where $\alpha$, $\beta$, and $\gamma$ are the functions of $\zeta$ and $\rho$ given by
\begin{align}
	\alpha &= \partial_\rho g^{\psi\psi}
	- \frac{\partial_\rho g^{\phi\phi}\partial_\zeta g^{\psi\psi}}{\partial_\zeta g^{\phi\phi}},
\\
	\beta &= -\partial_\rho g^{t\psi}
	+ \frac{\partial_\rho g^{\phi\phi}\partial_\zeta g^{t\psi}}{\partial_\zeta g^{\phi\phi}},
\\
	\gamma &= \partial_\rho g^{tt}
	- \frac{\partial_\rho g^{\phi\phi}\partial_\zeta g^{tt}}{\partial_\zeta g^{\phi\phi}}.
\end{align}
We can solve Eq.~\eqref{eq:lpsi} with respect to $l_\psi$ as the function of $\zeta$ and 
$\rho$ on $\Sigma$ in the form
\begin{equation}
	l_\psi^{\pm} = \frac{-\beta\pm\sqrt{\beta^2-\alpha\gamma}}{\alpha},
\label{sol:lpsi}
\end{equation}
and these lead to $l_\phi^\pm$ satisfying Eqs.~\eqref{eq:dU=0}, respectively. 
Therefore, there exist two branches of $\Sigma$,
$\Sigma^+ = \{(\zeta, \rho, l_\psi^+(\zeta, \rho), l_\phi^+(\zeta, \rho))\}$
and
$\Sigma^- = \{(\zeta, \rho, l_\psi^-(\zeta, \rho), l_\phi^-(\zeta, \rho))\}$.
They are projected uniquely into domains where $l_\psi^\pm(\zeta, \rho)$ and 
$ l_\phi^\pm(\zeta, \rho)$ take real values in the $\zeta$-$\rho$ plane.

\subsection{The case $\nu=0.1$}
\begin{figure}[!t]
\setlength\abovecaptionskip{0pt}
\begin{center}
\subfigure[]{
\includegraphics[width=80mm]{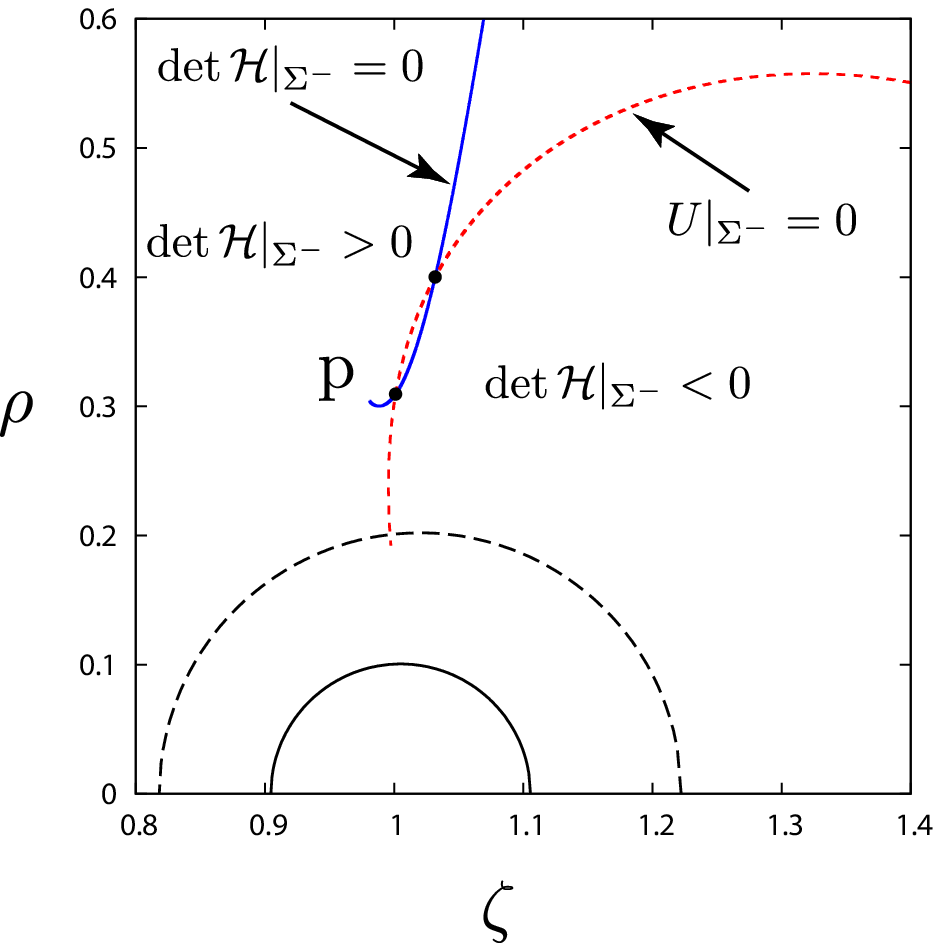}
\label{fig:real01}
}
\subfigure[]{
\includegraphics[width=74mm]{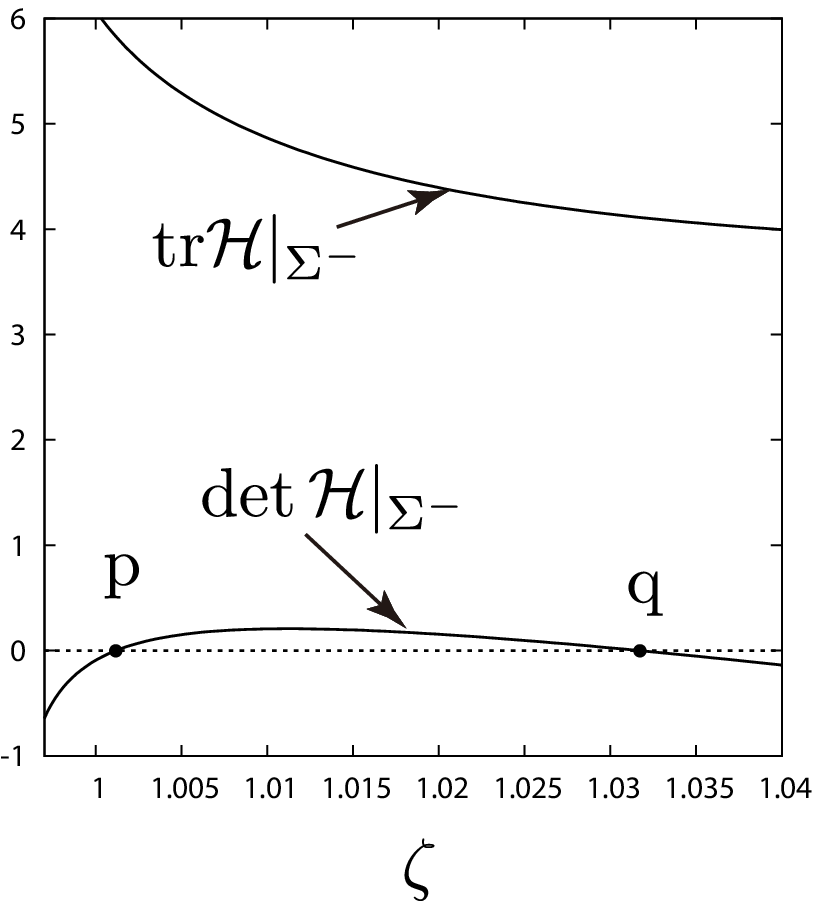}
\label{fig:real02}
}
\end{center}
\caption{\footnotesize
(a) The contours of $\det{\cal H}|_{\Sigma^-}=0$ ((blue) solid curve) and $U|_{\Sigma^-}=0$ 
((red) dashed curve) on $\Sigma^-$ are projected on the $\zeta$-$\rho$ plane in the case of 
the black ring with $\nu=0.1$. The solid half circle and the dashed half circle on the 
horizontal axis represent the event horizon and the ergosurface of the black ring, 
respectively.
(b) Plots of the value of $\det{\cal H}|_{\Sigma^-}$ and ${\rm tr}{\cal H}|_{\Sigma^-}$ 
as the functions of $\zeta$ along the contour $U|_{\Sigma^-}=0$. }
\bigskip
\end{figure}

We discuss the existence of stationary toroidal spiral orbits of massless particles in the 
black ring spacetime with $\nu=0.1$ as an example. We take $\Sigma^-$ first, and consider the 
contours $U|_{\Sigma^-}=0$ and $\det{\cal H}(U)|_{\Sigma^-}=0$ on it. The surface $\Sigma^-$ 
with the contours is uniquely projected into the $\zeta$-$\rho$ plane as shown 
in Fig.~\ref{fig:real01}. The contours $U|_{\Sigma^-}=0$ and $\det{\cal H}(U)|_{\Sigma^-}=0$ 
intersect at the points p and q, and the segment of $U|_{\Sigma^-}=0$ between p and q is in 
the region $\det {\cal H}(U)|_{\Sigma^-}>0$. We can check 
${\rm tr}~ {\cal H}(U)|_{\Sigma^-}>0$ on the segment separately as seen 
in Fig.~\ref{fig:real02}. As will be seen later, both of $l^-_\psi$ and $l^-_\phi$ are 
nonvanishing on the segment, and then stable toroidal spiral orbits of massless particles are 
realized at any points on the segment. The stable toroidal spiral orbits make a one-parameter 
family along the segment.

The stability condition is marginally satisfied at the points p and q. 
It means that there exist two marginally stable toroidal spiral orbits 
similar to the {\sl Innermost Stable Circular Orbit} (ISCO) of a massive particle 
moving around a black hole in four dimensions. The point p, near the event horizon, 
corresponds to the {\sl Innermost Stable Toroidal Spiral Orbit} (ISTSO) for a massless 
particle in the black ring spacetime, and the other point q is the 
{\sl Outermost Stable Toroidal Spiral Orbit} (OSTSO) for a massless particle 
in the black ring spacetime. By the same analysis, no stable toroidal spiral orbit 
appears in the other branch $\Sigma^+ $ for $\nu=0.1$. For thinner black rings, 
stable toroidal spiral orbits would be found in both the branches.

\subsection{Critical value of $\nu$}
We discuss the thickness parameter $\nu$ with which the black ring has the stable 
toroidal spiral orbits of massless particles. Numerical analysis shows 
that the crossing points p and q of contours $U|_{\Sigma^-}=0$ and 
$\det{\cal H}(U)|_{\Sigma^-}=0$ approach each other as $\nu$ increases. The points p and q 
merge together when $\nu$ takes a critical value $\nu_{\rm c}$, and the crossing points 
disappear for $\nu > \nu_{\rm c}$.

\begin{figure}[!t]
\setlength\abovecaptionskip{0pt}
\begin{center}
\subfigure[]{
\includegraphics[width=75mm]{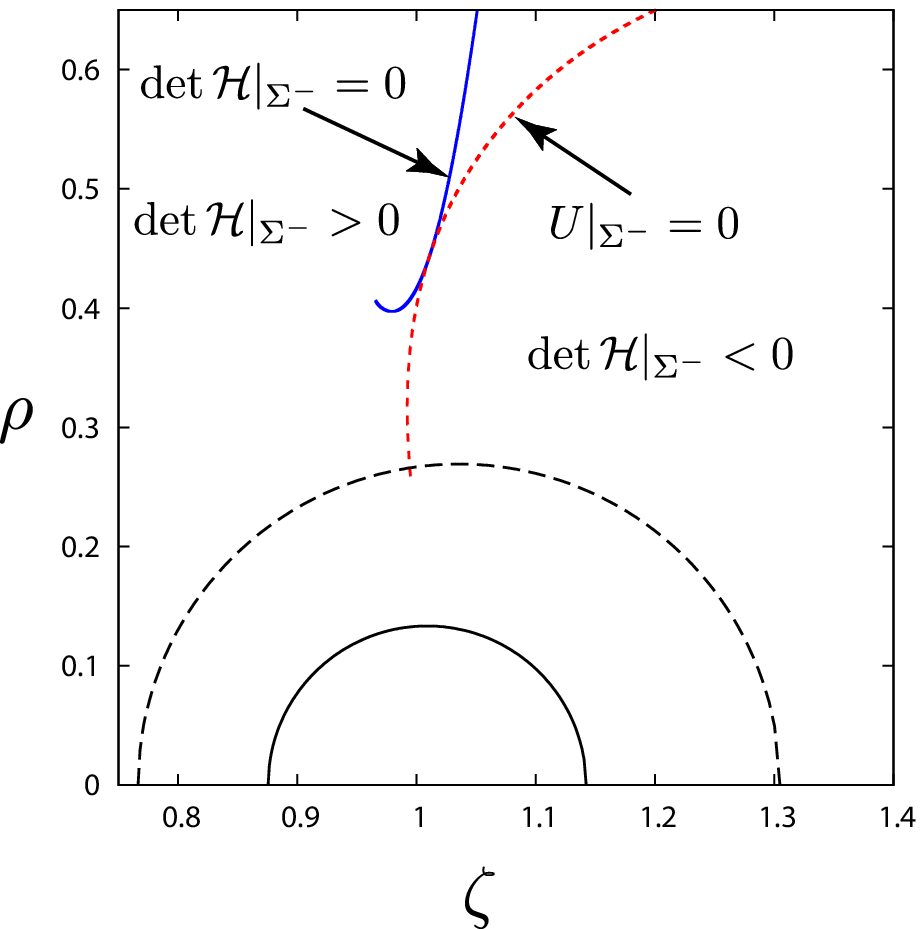}
\label{fig:real013224}
}
\subfigure[]{
\includegraphics[width=75mm]{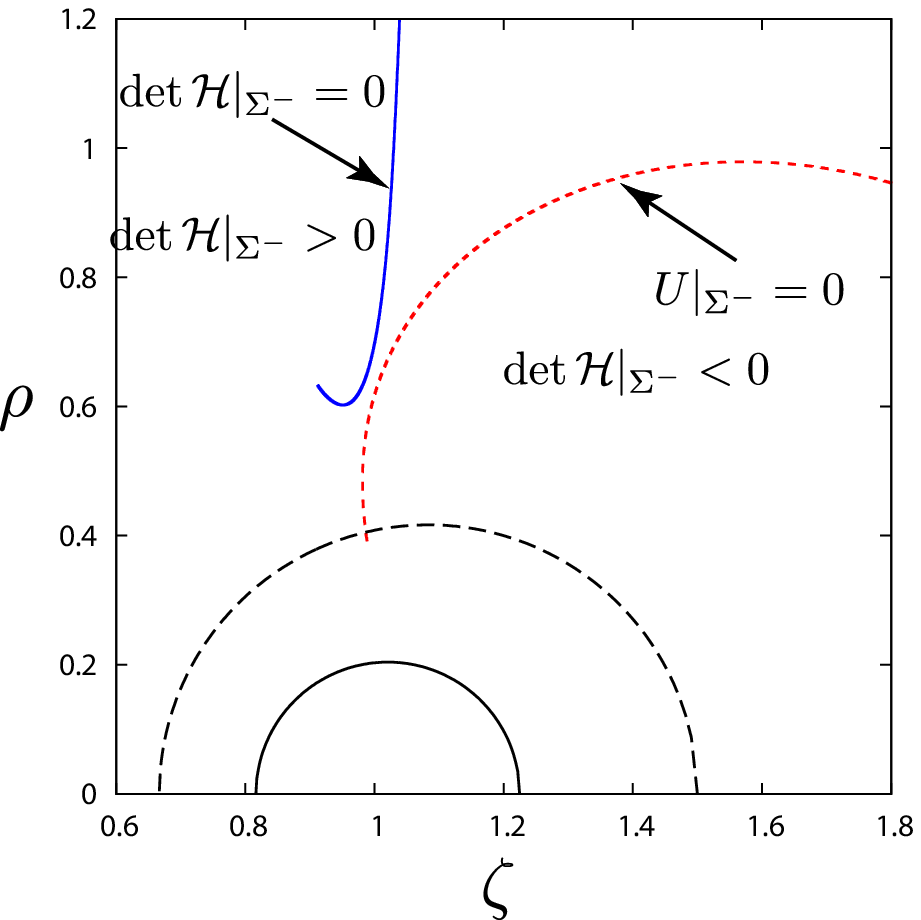}
\label{fig:real02}
}
\end{center}
\caption{\footnotesize
The (blue) solid curve and the (red) dashed curve in each panel are the projected contours of 
$\det{\cal H}|_{\Sigma^-}=0$ and $U|_{\Sigma^-}=0$ on the $\zeta$-$\rho$ plane, respectively, 
in the cases (a) $\nu=0.13224\simeq \nu_{\rm c}$ and (b) $\nu=0.2$. The solid half circle and 
the dashed half circle on the horizontal axis in each panel represent the event horizon and 
the ergosurface of the black ring, respectively.}
\label{fig:real02_crit}
\bigskip
\end{figure}

Figure~\ref{fig:real02_crit} shows the projection of $U|_{\Sigma^-}=0$ and 
$\det{\cal H}(U)|_{\Sigma^-}=0$ into the $\zeta$-$\rho$ plane for the cases 
(a) $\nu= 0.13224$ and (b) $\nu=0.2$. In the former case, the curve of $U|_{\Sigma^-}=0$ 
is tangent to the one of $\det{\cal H}|_{\Sigma^-}=0$ at a point, that is, the points p and q 
merge together. Hence, the critical value of $\nu$ is approximately given by 
$\nu_{\rm c}=0.13224\cdots$. The condition to determine 
the critical value $\nu_{\rm c}$ is discussed in Appendix~\ref{sec:appA}. On the other hand, 
the contours $U|_{\Sigma^-}=0$ and $\det{\cal H}(U)|_{\Sigma^-}=0$ have no 
crossing point in the case $\nu=0.2$. For the thin black rings with $\nu \leq \nu_{\rm c}$, 
there exist stable toroidal spiral orbits of massless particles, while for the black ring 
with $\nu>\nu_{\rm c}$ toroidal spiral orbits become unstable.

\section{Bounded orbits of massless particles}
\label{sec:4}
In this section, we consider nonstationary motion of massless particles trapped in a bounded 
domain outside the black ring horizon. It is natural to consider that there exist orbits 
wandering around the stable toroidal spiral orbits, which are discussed 
in the previous section. Such bounded orbits appear if particles are confined by an effective 
potential barrier. Since massless particles must satisfy the null condition~\eqref{eq:null2}, 
the contour $U=0$ is a set of turning points. Therefore, if a potential minimum with $U<0$ is 
surrounded by a closed contour $U=0$ outside the horizon, then bounded orbits exist. This is a 
sufficient condition for the existence of the bounded orbits.

\begin{figure}[!t]
\setlength\abovecaptionskip{0pt}
\begin{center}
\includegraphics[width=90mm]{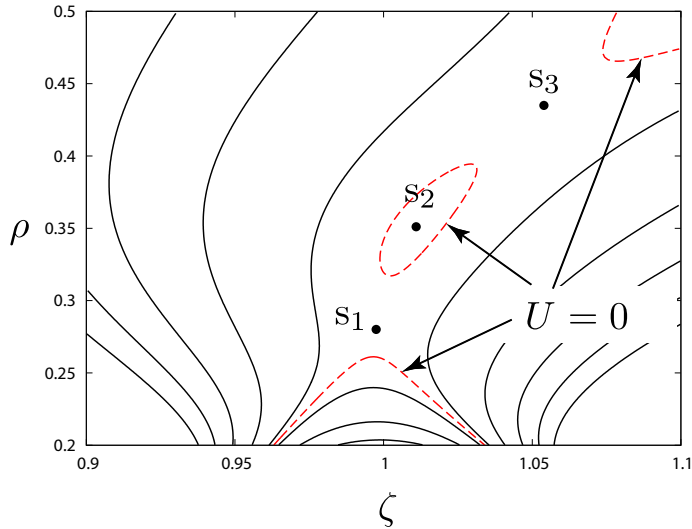}
\end{center}
\caption{\footnotesize
Contour plot of the effective potential $U$ with $l_\psi=-1.2651$ and $l_\phi=0.5204$ for 
$\nu=0.1$. The (red) dashed lines indicate the contour of $U=0$. 
The point s$_2$ is a local minimum, and s$_1$ and s$_3$ are saddle points.}
\label{fig:zero-pot}
\bigskip
\end{figure}

Figure~\ref{fig:zero-pot} shows contours of a typical $U$ that has a negative local minimum 
surrounded by the closed contour of $U=0$ in the case $\nu=0.1$. 
Massless particles are confined inside the potential barrier around a negative local 
minimum s$_2$. Indeed, we can verify that such a orbit is bounded by the closed contour $U=0$ 
by the numerical integration of the null geodesic equations~\eqref{eq:eom_x} and 
\eqref{eq:eom_k}.

\begin{figure}[!t]
\setlength\abovecaptionskip{0pt}
\begin{center}
\subfigure[]{
\includegraphics[width=75mm]{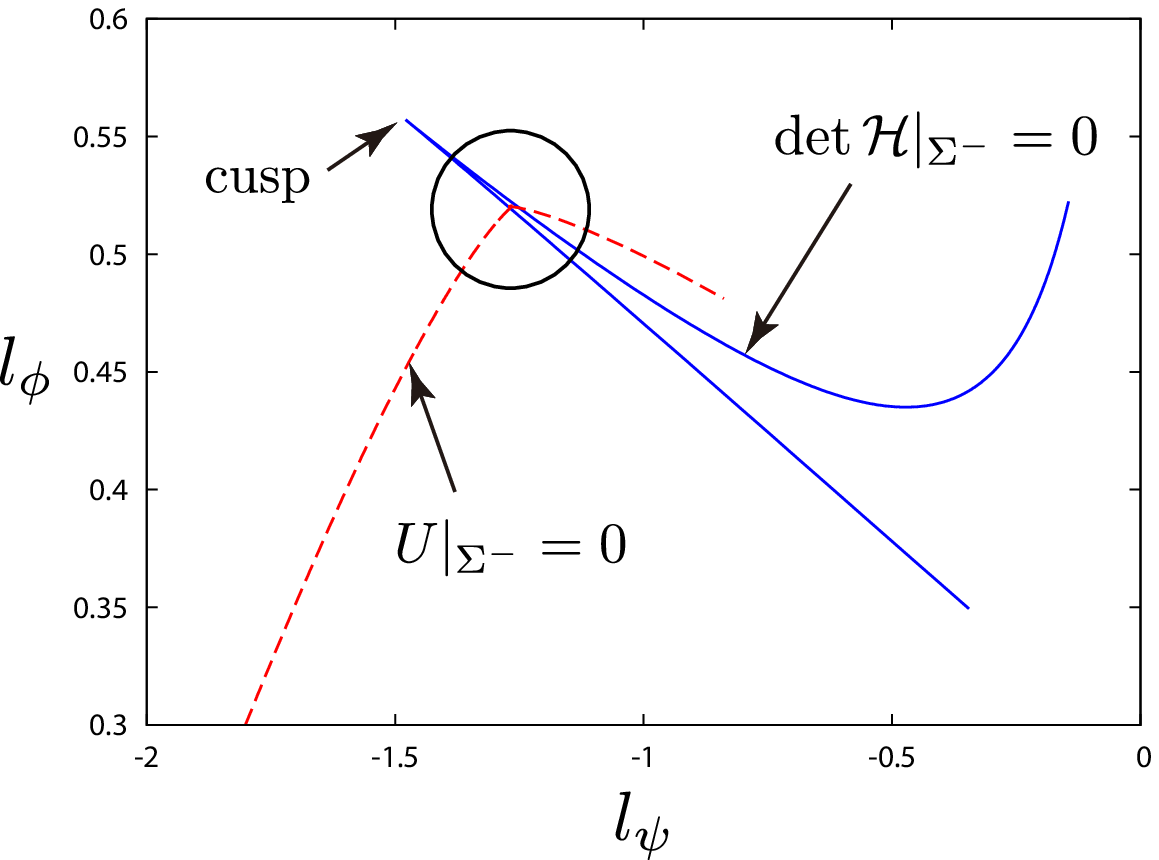}
\label{fig:phase01a}
}
\subfigure[]{
\includegraphics[width=75mm]{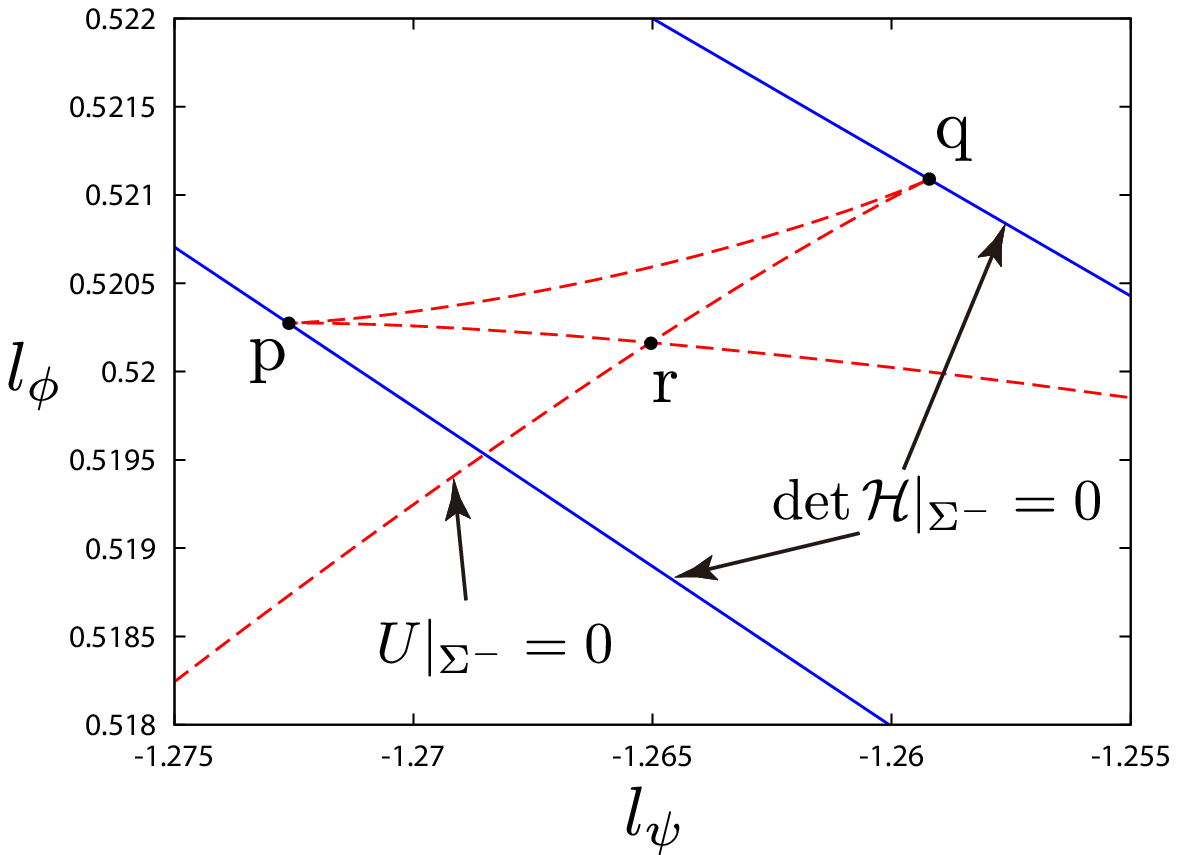}
\label{fig:phase01b}
}
\end{center}
\caption{\footnotesize
(a) Projection of the contours $\det{\cal H}|_{\Sigma^-}=0$ and $U|_{\Sigma^-}=0$ into the 
$l_\psi$-$l_\phi$ plane in the case $\nu=0.1$. The projection of the contour 
$\det{\cal H}|_{\Sigma^-}=0$ is shown by the (blue) solid curve, and the contour 
$U|_{\Sigma^-}=0$ is shown by the (red) dashed curve. There is a cusp on the projection 
of the contour $\det{\cal H}|_{\Sigma^-}=0$, and $\det{\cal H}|_{\Sigma^-}>0$ inside the wedge.
(b) A close-up view of the area inside the circle in the panel (a).}
\label{fig:phase01}
\bigskip
\end{figure}
\begin{figure}[!h]
\setlength\abovecaptionskip{0pt}
\bigskip
\begin{center}
\subfigure[]{
\includegraphics[width=80mm]{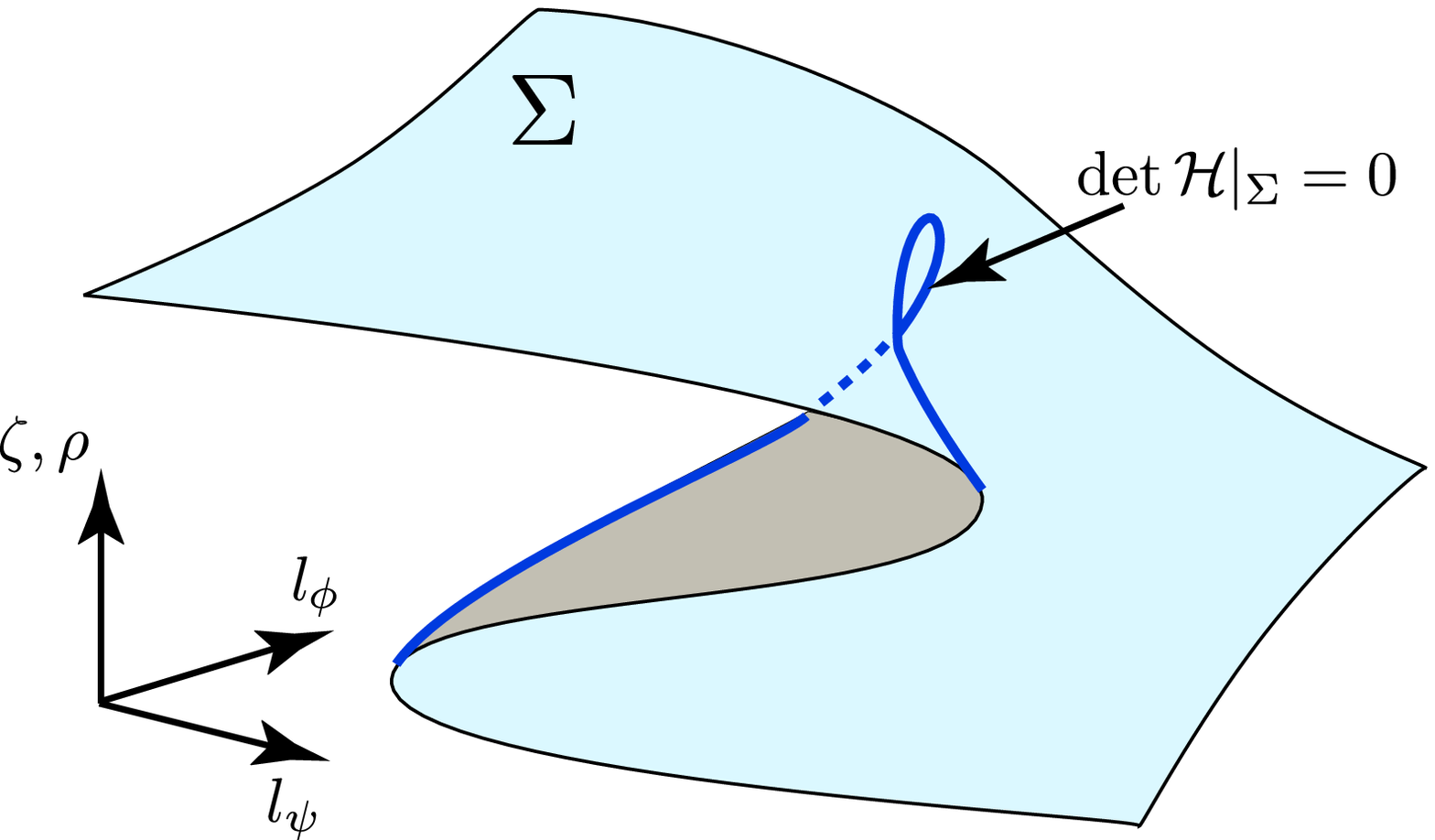}
}
\smallskip
\subfigure[]{
\includegraphics[width=180mm]{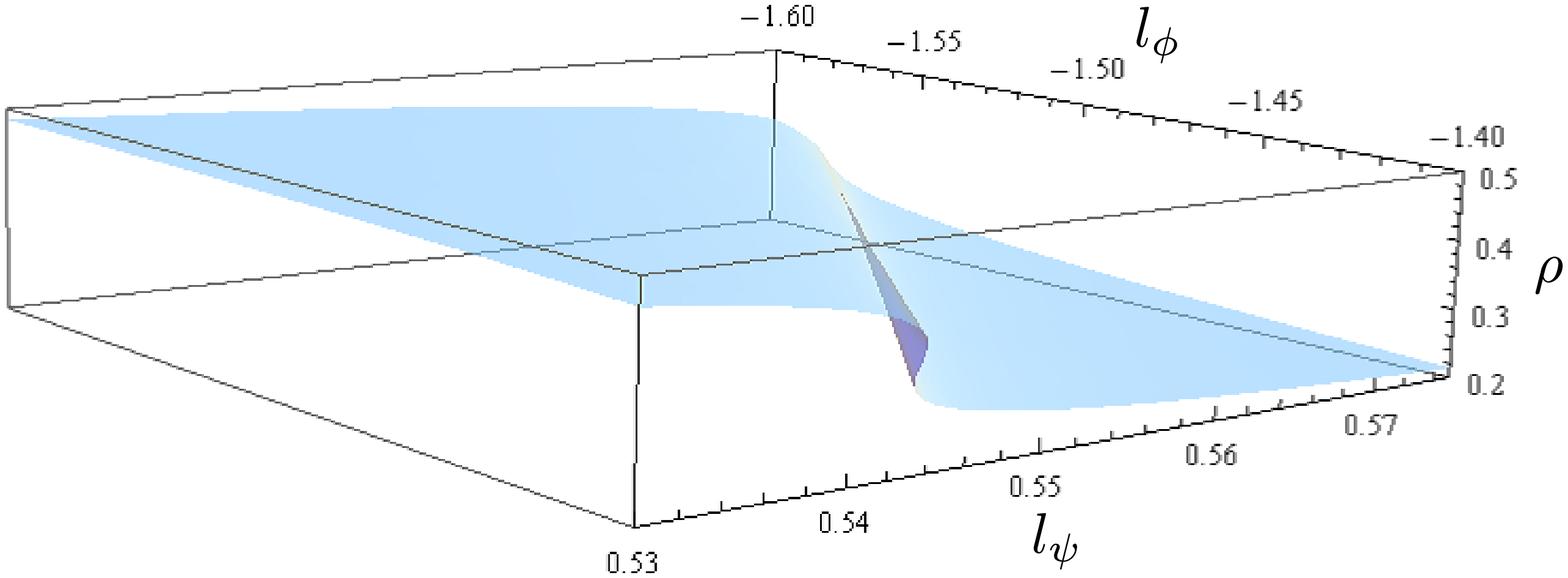}
\label{fig:s0}
}
\end{center}
\caption{\footnotesize
(a) Schematic diagram of the surface $\Sigma$. The contour of $\det{\cal H}=0$ is drawn 
by a solid curve.
(b) Numerical plot of $\Sigma^-$ in the three-dimensional space $(\rho, l_\psi, l_\phi)$ 
in the case $\nu=0.1$.
}
\label{fig:embedding_of_Sigma}
\end{figure}

\begin{table}[!t]
\setlength\abovecaptionskip{20pt}
\renewcommand\arraystretch{0.67}
\begin{tabular}{cccc}
\hline \hline
 ~~~~~~~~~~ & ~~ stationary points ~~&$~\det{\cal H}|_{\Sigma^-}$~~&~~$U|_{\Sigma^-}$~~\\
\hline
	t$_1$& s$_1$ & $-$ & $+$\\ \hline
	 & s$_1$ & $-$ & $+$\\
	t$_2$ & s$_2$ & $+$ & $-$\\
	 & s$_3$ & $-$ & $-$\\
\hline
	 & s$_1$ & $-$ & $+$\\
	t$_3$ & s$_2$ & $+$ & $-$\\
	 & s$_3$ & $-$ & $+$ \\
\hline
	 & s$_1$ & $-$ & $+$\\
	t$_4$ & s$_2$ & $+$ & $0$\\
	 & s$_3$ & $-$ & $+$\\
\hline
	 & s$_1$ & $-$ & $+$\\
	t$_5$ & s$_2$ & $+$ & $+$\\
	 & s$_3$ & $-$ & $+$\\
\hline
	t$_6$ & s$_3$ & $-$ & $+$\\
\hline\hline
\end{tabular}
\caption{
A variety of stationary points. The symbols $\pm$ show that 
${\rm det}{\cal H}|_{{\Sigma^-}}$ and $U|_{\Sigma^-}$ take a positive or a negative value, 
respectively.}
\label{stat_points}
\bigskip
\end{table}
Figure~\ref{fig:phase01} shows the projection of the contours $\det{\cal H}(U)|_{\Sigma^-}=0$ 
and $U|_{\Sigma^-}=0$ into the $l_\psi$-$l_\phi$ plane. A cusp of the projection of 
$\det {\cal H}(U)|_{\Sigma^-}=0$ appears (see Fig.~\ref{fig:phase01}(a)). The projection of 
$U|_{\Sigma^-}=0$ has a self-intersecting point, r, and two cusps, p and q, 
on the projection of $\det{\cal H}|_{\Sigma^-}=0$ (see Fig.~\ref{fig:phase01b}). 
These results mean that the embedding of $\Sigma^-$ is threefold. In fact, 
Figs.~\ref{fig:embedding_of_Sigma} show that a part of $\Sigma^-$ 
in the three-dimensional space of $(\rho, l_\psi, l_\phi)$ is threefold. 
The contour $\det{\cal H}(U)|_{\Sigma^-}=0$ makes the folded line, and hence the cusps 
and the crossing point on the $l_\psi$-$l_\phi$ plane are regular points on $\Sigma^-$.

\begin{figure}[!t]
\begin{minipage}[c]{.5\hsize}
\begin{center}
\includegraphics[width=80mm]{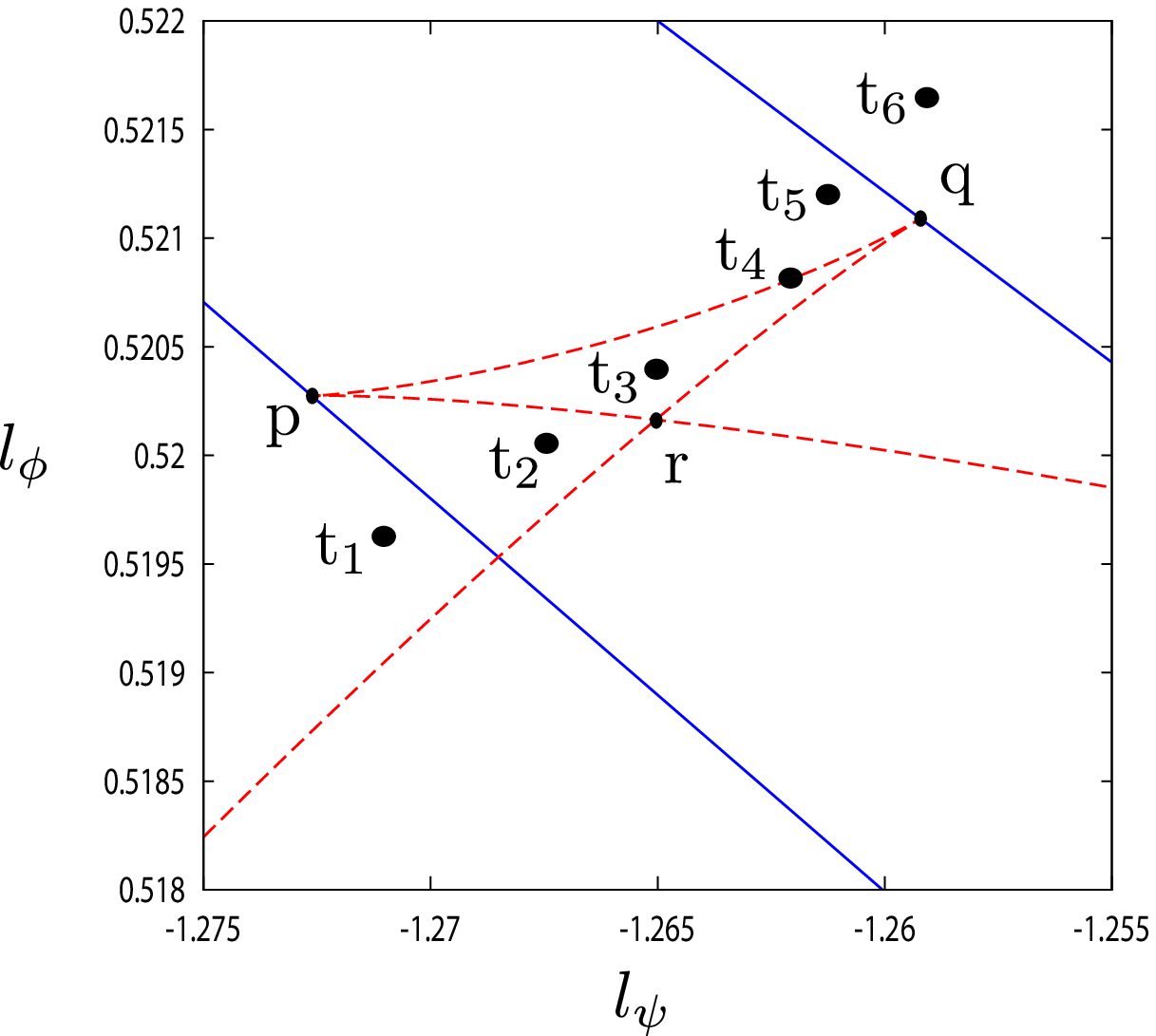}
\end{center}
\end{minipage}
\begin{minipage}[c]{.49\hsize}
\begin{center}
\includegraphics[width=30mm]{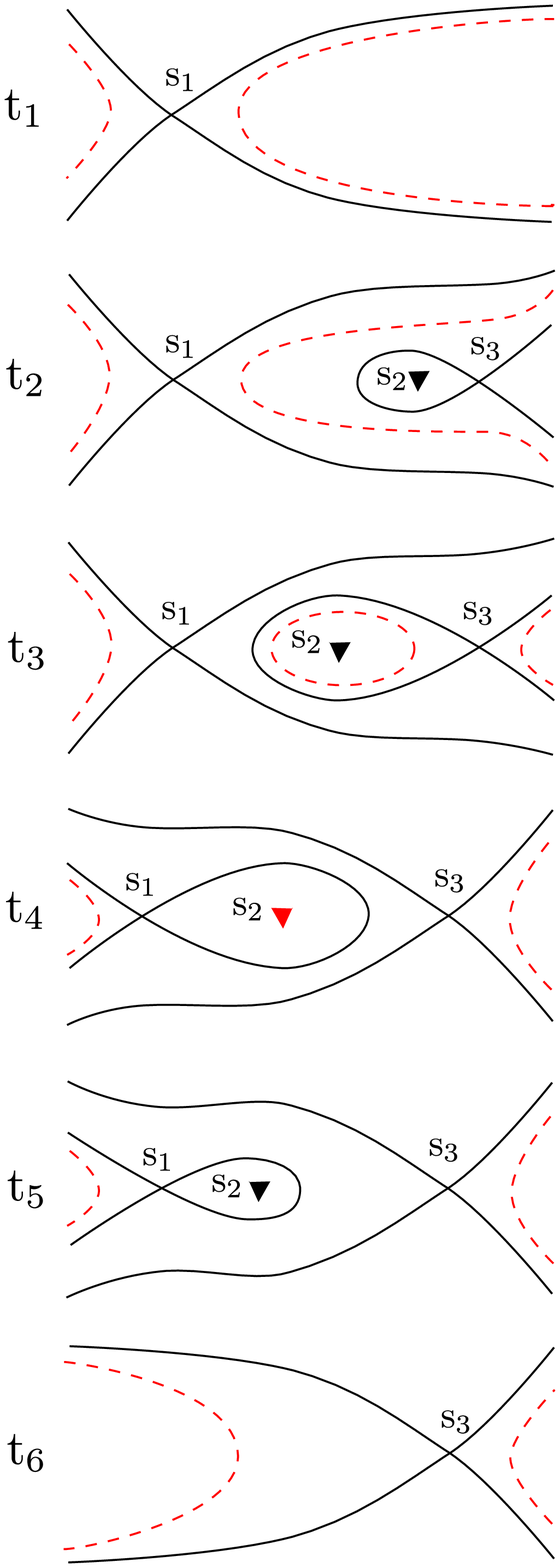}
\end{center}
\end{minipage}
\caption{\footnotesize
Schematic view of stationary points. Typical sets of parameters $(l_\psi, l_\phi)$ are shown 
by points t$_1$--t$_6$ in the $l_\psi$-$l_\phi$ plane (the left panel). 
The effective potentials corresponding to t$_1$--t$_6$ are shown by contour plots (the right 
panel). The solid lines are the contours of $U$ that are through saddle points, and the (red) 
dashed lines are the contours of $U=0$. The solid triangles show the local minimum points.}
\label{contours}
\bigskip
\end{figure}

In order to understand variety of the potential shapes that are dependent on the parameters 
$l_\psi$ and $l_\phi$, we consider several representative points, t$_1$--t$_6$, in the 
$l_\psi$-$l_\phi$ plane as shown in Fig.~\ref{contours}. For each set of the parameters 
$(l_\psi, l_\phi)$ in the onefold region of $\Sigma^-$, outside the wedge of 
$\det{\cal H}(U)|_{\Sigma^-}=0$, t$_1$ or t$_6$ for example, the stationary point of $U$ is 
a saddle point because $\det{\cal H}(U)|_{\Sigma^-}<0$ (see Table~\ref{stat_points}). 
On the other hand, for each set of the parameters $(l_\psi, l_\phi)$ in the threefold region, 
inside the wedge, t$_2$--t$_5$ for example, there exist three stationary points of $U$, 
two saddles and one local minimum. In particular, if we take a point in the triangle-like 
region qpr enclosed by the projection of $U|_{\Sigma^-}=0$, t$_3$ for example, 
$U<0$ at the local minimum, and $U>0$ at the two saddle points. In this case, 
the local minimum point is surrounded by a closed contour of $U=0$ (see Fig.~\ref{contours}), 
and then there exist bounded orbits of massless particles around the local minimum point.

A point on the projected segment of $U|_{\Sigma^-}=0$ between p and q, e.g., t$_4$, means 
that the effective potential admits stable toroidal spiral orbits, which are specified by a parameter, 
e.g., $l_\psi$, on the segment. A point on the segment between p and r or between q and r 
implies the potential well of $U$ from which massless particles can marginally leak out 
through the saddle point. The points p and q correspond to the ISTSO and the OSTSO, 
respectively. The effective potential at the point r has two marginally leaking saddle points 
with $U|_{\Sigma^-}=0$. In the case $\nu=\nu_{\rm c}$, the projection of 
$U|_{\Sigma^-}=0$ passes through the cusp of the projection of $\det{\cal H}(U)|_{\Sigma^-}=0$ 
on the $l_\psi$-$l_\phi$ plane, and then the triangle region qpr shrinks to a point. 
Thus, there is no stable bound orbit of massless particles for the black rings 
with $\nu>\nu_{\rm c}$.

\section{Summary}
We have investigated null geodesics in the singly rotating black ring geometry 
in five dimensions. As a result, we have found stable stationary orbits of massless particles 
in toroidal spiral shape if the thickness parameter $\nu$ is less than the critical value 
$\nu_{\rm c}= 0.13224\cdots$. The stable toroidal spiral orbits are a one-parameter family of 
the solutions characterized by a combination of the two nonvanishing angular momenta divided 
by the energy. As marginally stable orbits, there exist the innermost stable toroidal spiral 
orbit and the outermost stable toroidal spiral orbit. We have also shown the existence 
of nonstationary motion of massless particles trapped in a bounded domain outside the black 
ring horizon.

In four-dimensional black hole spacetimes, there exist stable stationary timelike geodesics 
like planetary orbits. In five dimensions, in contrast, spherical black holes with the asymptotic 
flatness seem unable to allow stable stationary orbits. However, the black rings with a 
thickness less than a critical value admit stable stationary timelike geodesics even though 
the spacetimes have the five-dimensional asymptotic flatness~\cite{Igata:2010ye, 
Igata:2010cd}. As for null geodesics around a black hole, it is well known that there exist 
unstable circular orbits of massless particles in four dimensions. The unstable circular 
orbits can be generalized in the higher-dimensional asymptotically flat spherical black hole 
cases. As far as we know, no stable bound null geodesic orbit around a gravitating body 
in a asymptotically flat spacetime was reported in any dimensions. 
The result in the present paper is the first example of the orbits of massless particles 
that are stably bound by the gravitational field outside a black object.

A fat black ring whose thickness is larger than the critical value $\nu_{\rm c}$ cannot bind 
any massless particles stably. Therefore, if the black ring that binds massless particles 
stably at an initial stage becomes fat with $\nu>\nu_{\rm c}$ 
by absorbing free-falling energy, then some part of the massless particles would be released 
toward infinity as radiation. This radiation would be a typical phenomenon for black rings, 
not for black holes. Furthermore, since the stationary points of the effective potential 
that allow stable bound orbits of massless particles are local minima, then massless particles 
trapped in a potential well outside the horizon can escape outward by quantum tunneling.

It is also an interesting question whether nonstationary bounded motion of massless particles 
in black ring spacetimes is chaotic or not. In the previous paper~\cite{Igata:2010cd}, 
we show an evidence of chaotic motion of massive particles by using Poincar\'e map. 
It suggests black rings allow no additional Killing tensor other than evidently known 
Killing vectors. Detailed investigations for null geodesics are important in relation to the 
conformal Killing tensor.

\subsection*{Acknowledgements}
This work is supported by Grant-in-Aid for JSPS No.J111000492 (T.I.) 
and Grant-in-Aid for Scientific Research No.19540305 (H.I.).

\appendix
\section{Conditions of the critical value $\nu_{\rm c}$}
\label{sec:appA}

As discussed in Sec.~\ref{sec:3}, there exist stable toroidal spiral orbits of massless 
particles, satisfying Eqs.~\eqref{eq:dU=0}, \eqref{eq:U=0}, and \eqref{eq:stability}, 
if $\nu$ is smaller than $\nu_{\rm c}$. As marginally stable toroidal spiral orbits, there exist 
the innermost stable toroidal spiral orbit (ISTSO) and the outermost stable toroidal spiral 
orbit (OSTSO), satisfying
\begin{equation}
	\left.\det{\cal H}(U)\right|_\Sigma=0.
\label{cond:marginal_stability}
\end{equation}
Hence, the Hessian matrix ${\cal H}(U)|_\Sigma$ has a zero-eigenvalue, or equivalently, 
the effective potential has a flat direction, i.e.,
\begin{equation}
	\partial_\sigma^2 U=0
\end{equation}
at the ISTSO and the OSTSO, where $\partial_\sigma$ is the eigenvector of 
${\cal H}(U)|_\Sigma$ associated with the zero-eigenvalue.

If $\nu$ increases to $\nu_{\rm c}$, the ISTSO and the OSTSO approach each other, 
and they degenerate in the limit $\nu=\nu_{\rm c}$. At the degenerate stationary point, 
the condition
\begin{equation}
	\partial_\sigma(\det{\cal H})=0,
\label{eq:degenerate}
\end{equation}
or equivalently,
\begin{equation}
	\partial_\sigma^3 U=0
\end{equation}
holds. The five conditions~\eqref{eq:dU=0}, \eqref{eq:U=0}, \eqref{cond:marginal_stability}, and 
\eqref{eq:degenerate} determine $\nu_{\rm c}$ and $(\zeta, \rho, l_\psi, l_\phi)$ for the degenerate point.

\end{document}